\documentclass[11pt,epsfig]{article}
\usepackage{graphicx}
\title{{\bf Stellar Populations and Chemical Evolution of
Late--Type Dwarf Galaxies}}
\author{Monica Tosi\\
\vspace{0.1cm}\\
\normalsize Osservatorio Astronomico, Via Ranzani 1, 40127 Bologna, Italy}
\date{}
\begin{document}
\maketitle
\def\bull{\vrule height .9ex width .8ex depth -.1ex}
\makeatletter
\def\ps@plain{\let\@mkboth\gobbletwo
\def\@oddhead{}\def\@oddfoot{\hfil\tiny
``Dwarf Galaxies and their Environment'';
International Conference in Bad Honnef, Germany, 23-27 January 2001}%
\def\@evenhead{}\let\@evenfoot\@oddfoot}
\makeatother

\begin{abstract}\noindent
Some aspects of the chemical evolution of late-type dwarf galaxies are 
reviewed, together with their implications on some issues of cosmological 
relevance. A more detailed approach to model their evolution is suggested.
The importance of deriving the star formation history in these systems 
by studying their resolved stellar populations is emphasized.
\end{abstract}

\section{Introduction} 
People have been studying the evolution of late-type dwarf galaxies for
two-three decades, but only in the last 5--6 years this has become
a fashionable research field. This late interest is
probably due to the circumstance that cosmologists have recently discovered
that understanding the evolution of the gaseous and stellar constituents
of these systems is a necessary step for a correct analysis of some
important problems of cosmological relevance. The connection between
late-type dwarfs and cosmology is at least three-fold: 
1) given the high gas content and low metallicity of irregular and blue
compact galaxies (hereinafter, Irrs and BCDs, respectively), they are the
closest analogues to primeval galaxies;
2) they are the systems currently providing the safest empirical estimate
of the primordial $^4$He abundance;
3) they have been suggested 5-6 years ago (e.g. Lilly et al. 1995, Babul 
\& Ferguson 1996) to be the local counterparts of the blue objects 
in excess in deep galaxy counts at intermediate redshifts.

Here I will try to summarize what can be learnt from chemical evolution
models and stellar population studies of Irrs and BCDs and what this
implies for these three issues of cosmological relevance. In the following
discussion, I will emphasize the open questions rather than the (important)
results already achieved in the field.

\section{Evolution of late-type dwarfs}

When a galaxy forms, after a while it begins to form stars and starts a 
series of cycles following in general the scheme drawn by Tinsley (1980). The
stars evolve and synthesize in their interiors heavier and heavier elements,
and then eject them in the surrounding medium when they lose mass
and die. In this way they pollute the interstellar medium (ISM) and modify
both its mass and chemical composition. In the meantime, the ISM mass and 
metallicity may change also for gas exchanges with adjacent regions (gas losses
or accretions, or both). The next generation of stars thus form in that 
ISM with a different initial composition and must have therefore a 
slightly different evolution. 

Models for galaxy chemical evolution take these phenomena into account, as 
well as their effects on the stellar and gaseous properties, by simplisticly
parametrizing the physical mechanisms. The major parameters are the star
formation (SF) law, the initial mass function (IMF), the gas flows in 
and out of the considered region, and the 
quantities involved in stellar nucleosynthesis (e.g. stellar lifetimes,
yields, mass loss, opacities, treatment of convection, etc.).

Chemical evolution models for Irrs and BCDs have been computed by several
groups in the last twenty years (e.g. Matteucci \& Chiosi 1983,
Matteucci \& Tosi 1985, Pilyugin 1993, Marconi et al. 1994, 
Carigi et al. 1995, Larsen et al. 2000, Recchi et al. 2001). The combination
of the predictions of these models with the information on the evolution of
Irrs and BCDs from other kinds of studies have interesting consequences on
the three issues of cosmological relevance mentioned in the Introduction.

\subsection{Irrs and BCDs as almost primeval galaxies ?}

Back in 1973, Searle et al. suggested that the blue colors, low
metallicities and high gas fractions of BCDs could be self-consistently 
explained only if these systems either are experiencing
now their first episode of SF activity or have a very discontinuous SF 
regime, dominated by short intense bursts separated by long quiescent phases
(see e.g. the top right-hand panel in Fig.\ref{sft}).

\begin{figure}
\vspace{5truecm}
\includegraphics{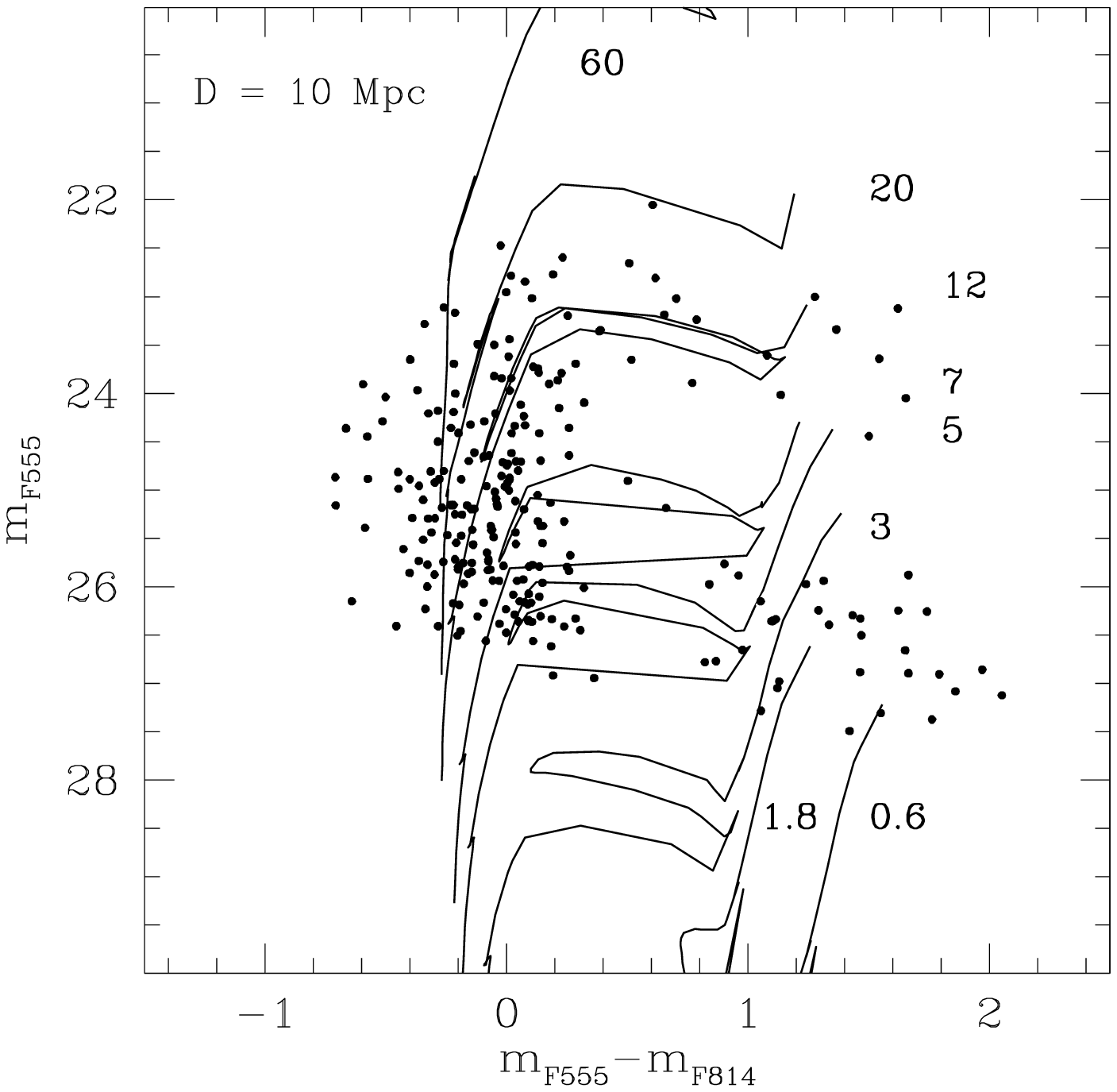}
\includegraphics{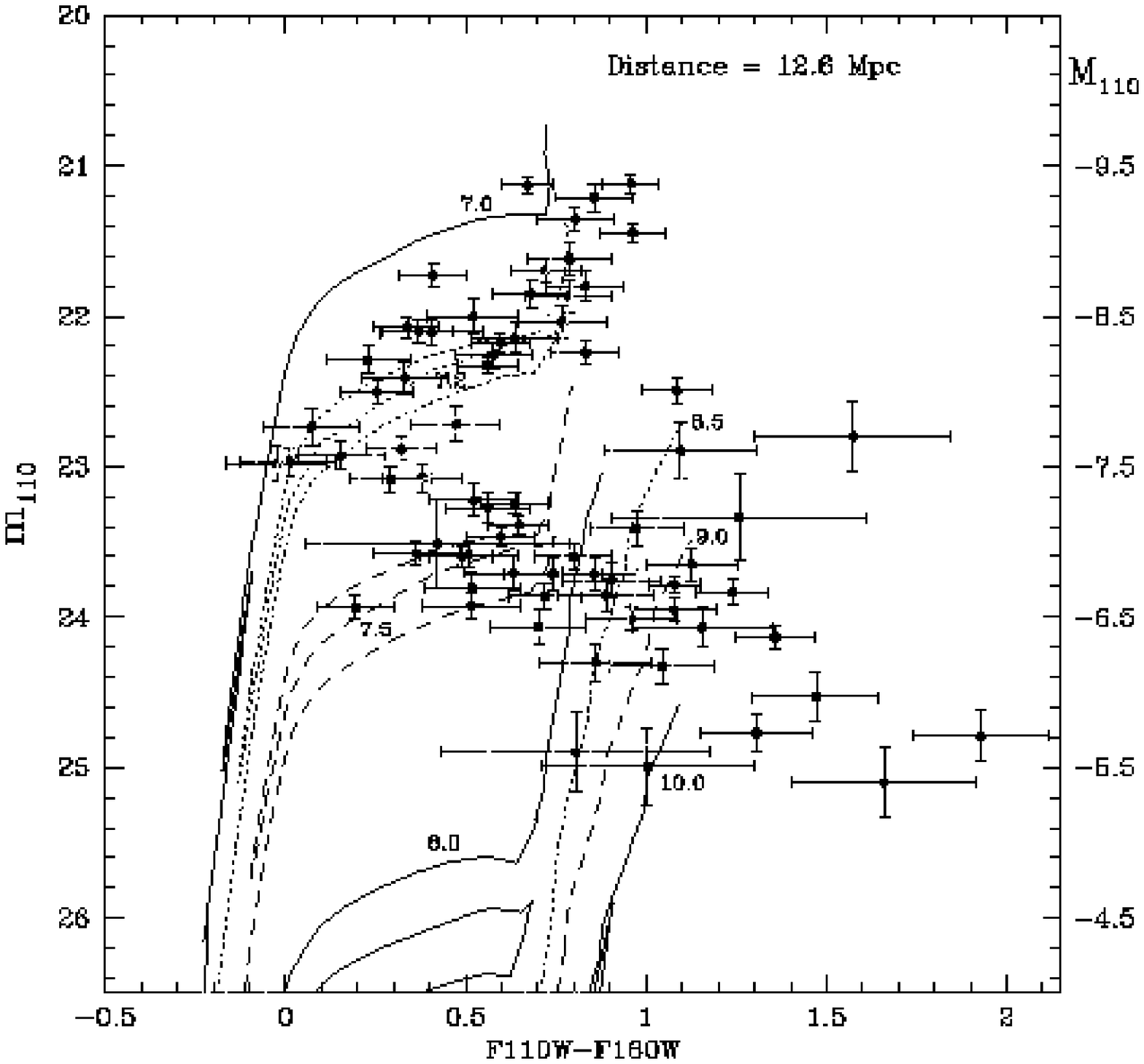}
\caption{Optical and NIR CMD of the resolved stars in the main body of
IZw18, as resulting from HST data (Aloisi et al. 1999 and Ostlin 2000,
respectively). Overimposed on the optical CMD are the Padova stellar evolution 
tracks with Z=0.0004 (masses are indicated in M$_{\odot}$). Overimposed
on the NIR CMD are the Geneva isochrones with the same metallicity.
\label{izw}}
\end{figure}

Almost thirty years later, however, no convincing case of galaxy at its
first SF experience has been found yet: all the systems for which 
sufficiently deep photometry exists reveal the presence of stellar populations
as old as the maximum lookback time allowed by the photometric depth
(see also Schulte-Ladbeck, this volume).
Even IZw18, the most metal-poor galaxy ever discovered, when imaged with
HST, has been recognized to contain stars at least a few hundreds Myr
old. Aloisi et al. (1999) have shown that the colour-magnitude diagram (CMD) 
derived from optical WFPC2 photometry of IZw18 contains several faint red
stars (see left panel in Fig.\ref{izw})
which can only be objects in the asymptotic giant branch (AGB) phase
of low and intermediate mass stars and must, therefore, have ages between
a minimum of 200-300 Myr up to a few Gyrs. Exactly the same conclusion has
been independently reached by Ostlin (2000) on the basis of near-infrared
Nicmos imaging (right panel in Fig.\ref{izw}).

Since lack of evidence is certainly not evidence of lack, this result does 
not exclude that really young galaxies do exist, but clearly restricts
their possible number to a small fraction of the known late-type dwarfs.
If Searle et al. (1973) were right, then, the vast majority of BCDs should
have a bursting mode of SF. This was indeed suggested by many chemical
evolution models (e.g. Matteucci \& Tosi 1985, Pilyugin 1993), which confirmed
that the best way to reproduce simultaneously colours, chemical abundances
and gas fractions of BCDs (and possibly of Irrs) was to have only a few
(7--10, at most) episodes of strong SF activity, with long quiescent intervals.
This conclusion has however been questioned by more recent models (e.g. Carigi
et al. 1995, Legrand 2000) suggesting that continuous SF regimes, with
very low rates, could account equally well for the observed features.

The lack of unique scenarios from chemical evolution models of late-type
dwarfs is due to the fact that we do not have yet sufficient observational 
data to properly constrain them. For what concerns the SF regime, the
most direct way to infer it from observations is from the CMDs of systems
close enough to be resolvable in single stars.

\subsection{The helium--metallicity relation}

The primordial $^4$He abundance is nowadays one of the hottest issues in
cosmology, due to the recent observational data (Tytler et al. 2000 and 
references therein) showing that primordial D was quite low,
(D/H)$_p\simeq3\times10^{-5}$, and, hence, either $^4$He was high (in mass
fraction Y$_p>$0.24), or the standard theories of Big Bang Nucleosynthesis 
are not self-consistent. The only method currently considered reliable to
derive the primordial helium abundance is based on the extrapolation to
zero-metallicity of the helium--metallicity relation derived from observations
of HII regions in metal-poor galaxies. Since the heavy elements more reliably
derived from HII regions are oxygen and nitrogen, what people actually use 
are the helium--oxygen and helium--nitrogen relations, inferred
from a linear fit to the data (see e.g. Izotov et al. 1997).

\begin{figure}
\vspace{4truecm}
\includegraphics{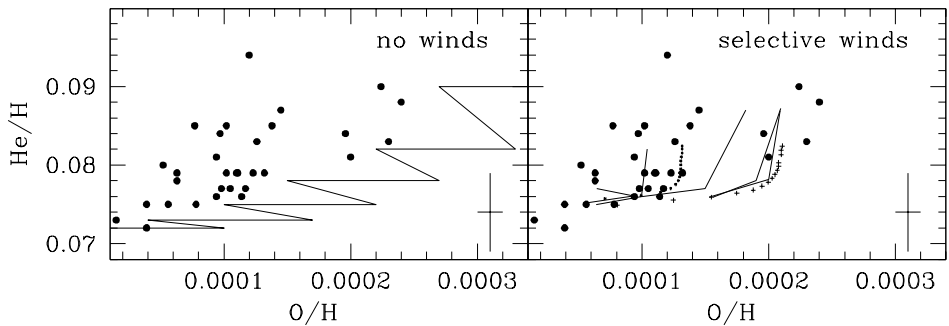}
\caption{Distribution of the helium vs oxygen abundance as derived from HII
regions (dots) in late-type dwarfs and as predicted (curves) by chemical 
evolution models
(Pilyugin 1993 on the left panel, Marconi et al. 1994 on the right panel).
\label{chemev}}
\end{figure}

One of the questions on this method is whether or not these relations can
be taken as linear. He, N and O are synthesized in different
stellar mass ranges: helium is produced by all stars, nitrogen almost
exclusively by intermediate mass stars, and oxygen only by massive ones;
and this implies that the timescales for their ISM enrichment are quite
different from each other. This point was explicitely addressed by Pilyugin
(1993) who showed with his chemical evolution models that the evolution 
of He with O (and with N) in a BCD is not linear but has typically
a saw-tooth shape (see left-hand panel in Fig.\ref{chemev}).

Furthermore, all chemical evolution models of these galaxies invoke the
occurrence of significant galactic winds during their life,
and these winds are often required to be more effective in removing elements
(like oxygen) expelled from massive stars. If these differential winds
occur, this also makes the behaviour of helium with oxygen non linear during
the lifetime of a dwarf, as shown both by Pilyugin (1993) and Marconi et al.
(1994) and illustrated in the right-hand panel of Fig.\ref{chemev}. 
The questions are: Do the winds actually occur and in what percentage
of Irrs and BCDs ? If they do occur, do they really remove some (which~?)
elements for ever ? With what efficiency~?
Several optical and X-ray observations of ionized filaments and bubbles in
some Irrs and BCDs confirm that somewhere the phenomenon does take place
and seem to preferentially eject alpha-elements (see e.g. Meurer et al. 1992,
Sahu \& Blades 1997, Hensler et al. 1998, for NGC1705). However, hydrodynamical
models of the SN ejecta in dwarf galaxies do not
agree yet on the timescales or on the efficiency of the gas removal (cfr.
Silich \& Tenorio-Tagle 1998, D'Ercole \& Brighenti 1999, Mac Low \& Ferrara
1999).

In spite of the above arguments, I personally believe that a linear fit
to the observed helium--oxygen and helium--nitrogen distributions derived
from HII regions is the correct approach to infer the primordial $^4$He. 
In fact, each data point in the empirical distribution corresponds to the 
final (i.e. present epoch) stage of the He--O and He--N behaviours
in each galaxy, and to derive the mean initial $^4$He one needs to average
over all the possible behaviours of the whole sample of galaxies. The
observed shape of the distribution suggests indeed that this is well 
represented by a linear fit.

\subsection{Local counterparts of faint blue galaxies ?}

Some authors (e.g. Lilly et al. 1995, Babul \& Ferguson 1996) have suggested
that starbursting dwarfs may be the local counterparts of the faint blue
galaxies usually found in excess in deep galaxy counts at redshift of
0.7--1.0. To verify whether this is a likely possibility, one needs to
check if most of the dwarfs have been forming stars at the epochs corresponding
to that redshift (hence about 6--7 Gyr ago) and if, in case, that SF activity
was strong enough (i.e. $\sim$1 M$_{\odot}$yr$^{-1}$)
to guarantee the observed luminosity levels.

We have already mentioned above that all the galaxies examined so far 
appear to contain stars as old as the lookback time corresponding to
the faint mag limit of the available photometry. This makes it extremely 
probable that most of them have already been active several billion years 
ago. The tough question is on the intensity of the SF.
Van Zee (2001) has very recently surveyed the SF histories of a large sample
of distant, isolated dwarf Irrs and concludes that their SF activities are
too modest and it is very unlikely that these systems could significantly
contribute to the faint blue galaxy excess. 

This conclusion is inevitably based on integrated properties of these
distant objects. To obtain a more quantitative information, it is necessary
to examine nearby systems where the SF history can be derived directly from
the resolved stellar populations and a more detailed evaluation of the
SF rates at different epochs can be performed.

\section{SF histories from stellar populations}

\begin{figure}
\vspace{10.5truecm}
\includegraphics{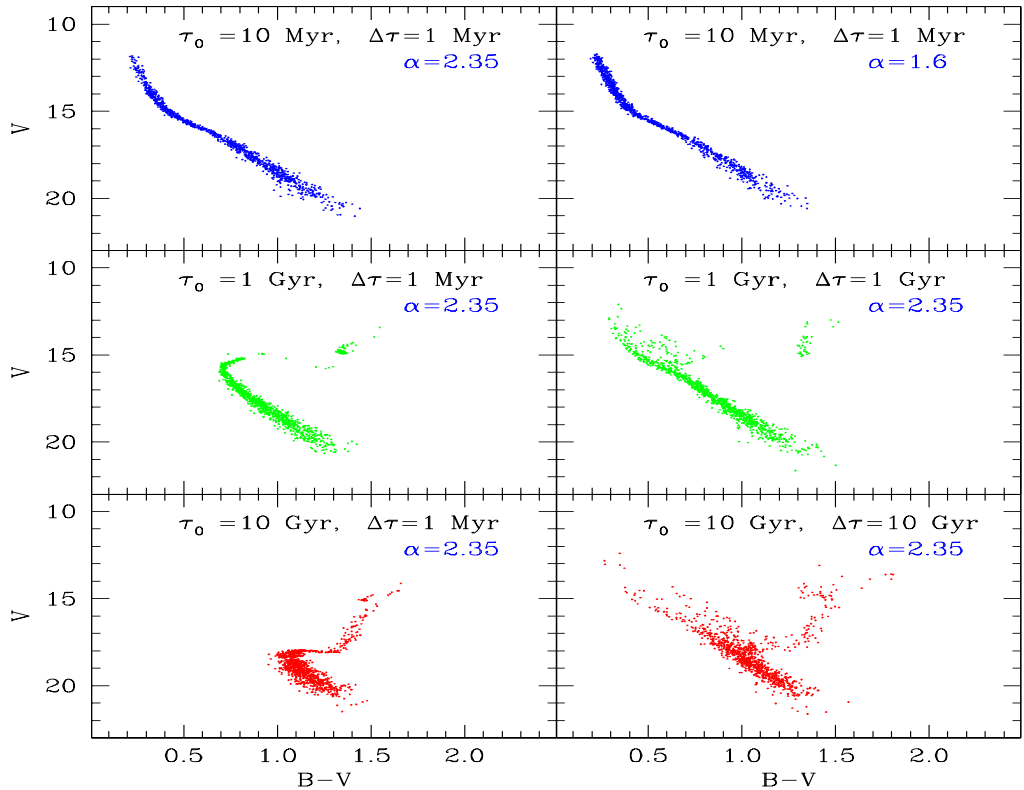}
\caption{Synthetic CMD of a hypothetical stellar system with 1000 single 
stars photometrically resolved. Distance modulus, (m-M)$_0$=12.5, reddening,
$E(B-V)$=0.45, photometric errors and incompleteness are those derived from 
observations of a 
real Galactic open cluster (NGC2660). The adopted stellar evolution tracks
(Bressan et al. 1993) have solar metallicity. The assumed starting time
(backwards from present epoch) and duration of the SF activity are labelled 
in each panel, together with the adopted IMF slope.
\label{cmd}}
\end{figure}

As discussed in the previous sections, for a correct approach to study the
evolution of Irrs and BCDs and for several issues of cosmological relevance,
it is of crucial importance to derive the SF history of a large number of
galaxies of these types. The most direct estimate of the epochs and of
the intensities of the SF activities is derivable from the CMDs, since their
morpholoy is the consequence of the system evolutionary conditions. 
Fig.\ref{cmd} shows a few examples of how the CMD morphology is affected by 
the SF history and IMF. We thus proposed several years ago 
(Tosi et al. 1989, 1991) the method of synthetic CMDs to
derive from the observational CMDs as much information as possible,
taking into account all the incertainties related to both observations
and theories. Similar procedures have been developed also by other groups
(e.g. Aparicio et al. 1996, Tolstoy \& Saha 1996) and are now widely applied
(see also the contributions by Grebel, Schulte-Ladbeck, Tolstoy, this volume).

The numerical procedure creates synthetic CMDs, via MonteCarlo extractions 
on homogeneous sets of stellar evolution tracks (e.g. the Geneva or the 
Padova sets) of various metallicities, taking into account all the 
involved theoretical parameters and uncertainties (age, metallicity, IMF,
SF law, stochastic effects due to small number statistics, etc.). The
synthetic CMDs have the observed number of stars of the examined region
and its observational uncertainties (photometric errors, incompleteness
factors, blending). The latter is a very important point which requires
a special accuracy during the photometric analysis.
A model can be considered satisfactory only
if it reproduces all the features of the observational CMD and luminosity
function (LF). 

By comparing all the simulated cases with the empirical CMD and LF we
select those which are more consistent with them and thus derive the
number, starting epoch, duration and intensity of the SF episodes, their
IMF, the durations of significant quiescent intervals (if any), and hints
on the most likely metallicity of the various stellar generations.
In most, if not all, the cases, the results are not unique, but we can
sensibly reduce the range of acceptable values of the various parameters,
thus obtaining a good indication on the most likely evolution of the
examined region.

\begin{figure}
\vspace{5truecm}
\includegraphics{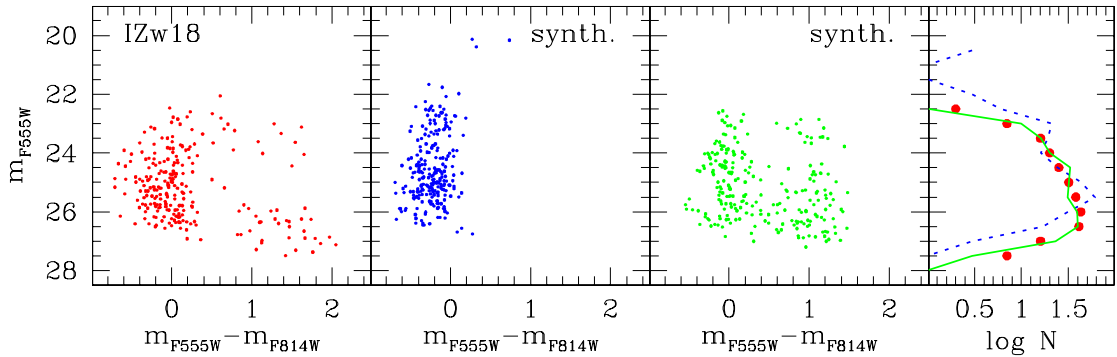}
\caption{The case of IZw18. The left panel shows the empirical V,V--I CMD 
(in the HST--Vegamag system) derived from HST--WFPC2 photometry. The two
middle panels show two cases of synthetic CMDs with the same number of
objects, photometric error and incompleteness factors as in the empirical
CMD. The left hand case refers to one single burst of SF from 10 Myr ago
to the present epoch, with a steep IMF slope $\alpha$=3.0. The right hand
case assumes two SF episodes: one from 1 Gyr to 30 Myr ago and the other
from 20 Myr to 5 Myr ago, with a flat IMF slope $\alpha$=1.5. Finally, 
the right-hand panel presents the comparison of the LFs corresponding to the
two synthetic cases (dotted line for the single episode and solid line for
the two episodes) with the empirical one (dots). See Aloisi et al. (1999) for
further details.
\label{izwsyn}}
\end{figure}

We have applied the method to Local Group Irrs with ground-based observations
and to more distant late-type dwarfs observed with HST (see Table 1 for
a summary of the results). Fig.\ref{izwsyn} shows two cases of synthetic
CMDs for IZw18: it is apparent that a single recent SF burst cannot account
for the observed red stars (both faint and bright) and predicts too many
too bright objects, despite the rather steep IMF. The case with two episodes
is instead consistent with the observed features.
The most recent application is being performed
on the BCD NGC1705 and its preliminary results are described by Annibali 
et al. (this volume). Of the 8 dwarfs examined so far by our group, only
NGC1569 shows evidence for a very intense and short burst (Greggio et al. 
1998), in addition to
other longer, less striking episodes. All the others appear to have had a 
gasping SF regime, with episodes of much more moderate activity. None
of them shows any evidence for quiescent phases longer than 10$^8$ yr.

\begin{figure}
\vspace{12.5truecm}
\includegraphics{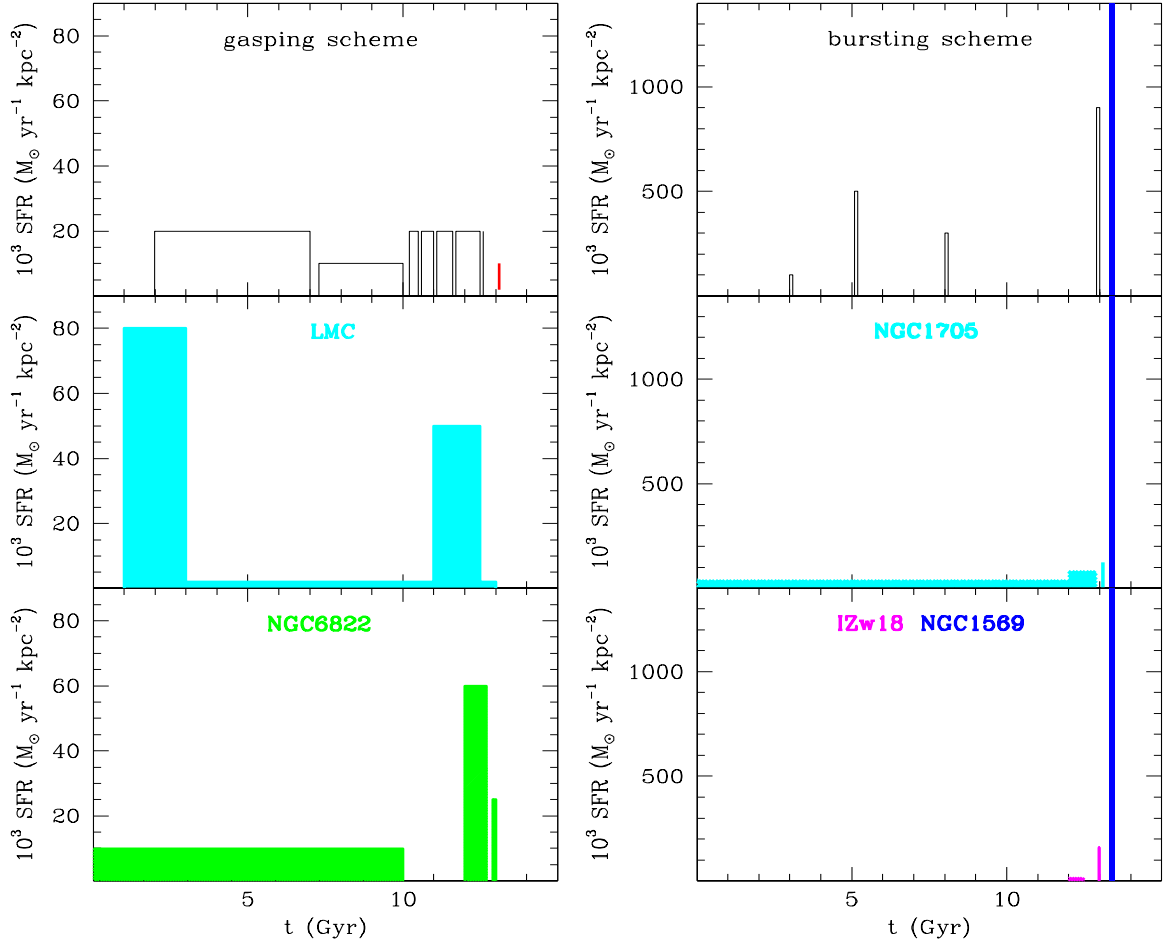}
\caption{SF histories in late--type dwarfs. The hypothetical time behaviour 
of the SF rate per unit area is sketched in the top panels for the gasping 
(left) and bursting (right)
regimes. The middle and bottom panels show the actual behaviours
derived for 5 galaxies: LMC (Pilyugin 1996, Pagel \& Tautvaisiene
1998), NGC1569 (Greggio et al. 1998), NGC1705 (Annibali et al. this volume),
NGC6822 (Marconi et al. 1995, Gallart et al. 1996), IZw18 (Aloisi et al. 1999).
The long spike crossing all the right-hand panels shows the striking case of
NGC1569, which has a SFR per unit area of 4 M$_{\odot}$yr$^{-1}$kpc$^{-2}$.
\label{sft}}
\end{figure}

If we combine the results on the SF history of the late-type dwarfs studied
by our group with those obtained by other groups (see e.g. Grebel 1998 for
a comprehensive review of Local Group galaxies and Schulte-Ladbeck,
this volume, for BCDs outside the Local Group) it is apparent that these
galaxies, with very few exceptions, have had qualitatively similar SF
histories. Fig.\ref{sft} shows the SF histories derived in 5 galaxies: NGC1569
is the only one which appears to follow the bursting scheme. The data available
so far have not allowed yet to accurately derive its earlier activity, but
the new HST photometry in the near infrared by Aloisi et al. (2001) should
allow for a lookback time of, at least, a few Gyrs. In summary:
\begin{itemize}
\item
Both Irrs and BCDs seem to have a roughly continuous or gasping
SF regime, with various episodes of moderate activity and no clear sign
of long periods of total inactivity. 
\item
Only the few cases of the type of NGC1569 have bursts with SF rates high 
enough to allow for the luminosity attributed to faint blue galaxies 
at redshift 0.7--1.
\item
No evidence has been found yet for dwarfs at their first SF activity.
\end{itemize}

{\small
\begin{table*}
\begin{center}
\caption{Dwarf irregulars and BCDs in our program. Listed for each
galaxy are: name, coordinates, derived distance 
modulus, number of resolved stars with small photometric error, and SF mode
derived from the synthetic CMD method.}
\begin{tabular}{lccccc}
\hline\hline
Name & R.A. & DEC&  (m$-$M)$_o$ & sel obj & SF mode\\
\hline
DDO ~~70 (Sex B)   &09 57 23&+05 34 07& 25.6 &1300 & gasps \\
DDO 209 (NGC 6822) &19 42 07&$-$14 55 01& 23.5 &1772& gasps \\
DDO 210 (Aquarius) &20 44 08&$-$13 02 00& 28 ?& 633& gasps \\
DDO 221 (WLM)      &23 59 23&$-$15 44 06& 25.0 & 2000& gasps \\
DDO 236 (NGC 3109) &10 00 48&$-$25 55 00& 25.7 &2605& gasps \\
\hline 
NGC 1569    & 04 26 04 & +64 44 29   & 26.7 & 801& gasps+burst \\
IZw18       & 09 34 02 & +55 14 19   & 30.0 & 150& gasps \\
NGC 1705    & 04 54 13 & $-$53 21 40 & 28.6 & $17000$& gasps \\
\hline 
\end{tabular}
\end{center}
\end{table*}
}

\section{Future prospects}

From the above presentation, it is apparent that further studies on several
issues are still required to obtain a reliable scenario for the evolution
of late-type dwarfs. In particular, the SF histories and the major 
characteristics of the gas flows triggered by SN explosions should be
examined in more details. For this reason I believe that, at present, 
the correct approach to model the chemical evolution of these systems would
be to concentrate on single representative cases, rather than to model the
overall features of a large sample of galaxies. For the chosen cases I would:
 \begin{itemize}
\item
Try to get as much information as possible on the evolutionary parameters
from observations (e.g. IMF and SF from populations synthesis and synthetic
CMD; gas distribution and flows from multiwavelength observations; etc.);
\item Compute the hydrodynamics of the ISM and SN ejecta in the conditions
of the examined galaxy derived from the first point;
\item Only at this point, model the chemical evolution of the examined
galaxy including all the results of the previous points as input data.
\end{itemize}

\noindent
{\bf Acknowledgements}

\noindent
I warmly thank the organizers of this enjoyable and interesting meeting
for the invitation and the financial support. I sincerely congratulate them 
for the success of their Graduate School. This work has been partly supported 
by the Italian MURST through Cofin2000.

{\small
\begin{description}{} \itemsep=0pt \parsep=0pt \parskip=0pt \labelsep=0pt
\item {\bf References}

\item
Aloisi, A., Tosi, M., Greggio, L. 1999, AJ 118, 302

\item
Aloisi, A., Clampin, M., Diolaiti, E., Greggio, L., Leitherer, C., Nota, A.,
 Origlia, L., Parmeggiani, G., Tosi, M. 2001, AJ 121, 1425

\item
Aparicio, A., Gallart, C., Chiosi, C., Bertelli, G. 1996, ApJ 469, L97

\item
Babul, A. \& Ferguson, H.C. 1996, ApJ 458, 100

\item
Bressan, A., Fagotto, F., Bertelli, G., Chiosi, C. 1993, A\&AS 100, 647

\item
Carigi, L., Colin, P., Peimbert, M., Sarmiento, A. 1995, ApJ 445, 98

\item
D'Ercole, A. \& Brighenti, F. 1999, MNRAS 309, 941

\item
Gallart, C., Aparicio, A., Bertelli, G., Chiosi, C. 1996, AJ 112, 1950

\item
Grebel, E.K. 1998, IAU Symp.192, p.1

\item
Greggio, L., Tosi, M., Clampin, M., De Marchi, G., Leitherer, C., Nota,
 A., Sirianni, M. 1998, ApJ 504, 725

\item
Hensler, G., Dickow, R., Junkes, N., Gallagher, J.S. 1998, ApJ 502, L17

\item
Izotov, Y.I., Thuan, T.X., Lipovetsky, V.A. 1997, ApJS 108, 1

\item
Larsen, T.I., Sommer-Larsen, J., Pagel, B.E.J. 2000, astro-ph/0005249

\item
Lilly, S.J., Tresse, L., Hammer, F., Crampton, D., Le Fevre, O. 1995, ApJ
 455, 108

\item
Mac Low, M.-M., \& Ferrara, A. 1999, ApJ 513, 142

\item
Marconi, G., Tosi, M., Greggio, L., Focardi, P. 1995 AJ 109, 173

\item
Marconi, G., Matteucci, F., Tosi, M. 1994, MNRAS 270, 35

\item
Matteucci, F., \& Chiosi, C. 1983, A\&A 123, 121

\item
Matteucci, F., \& Tosi, M. 1985, MNRAS 217, 391

\item
Meurer, G.R., Freeman, K.C., Dopita, M.A., Cacciari, C. 1992, AJ 103, 60

\item
Ostlin, G. 2000, ApJ 535, L99

\item
Pagel, B.E.J. \& Tautvaisiene, G. 1998, MNRAS 299, 535

\item
Pilyugin, L.S. 1993, A\&A 277, 42

\item
Pilyugin, L.S. 1996, A\&A 310, 751

\item
Recchi, S., Matteucci, F., D'Ercole, A. 2001, MNRAS in press 

\item
Sahu, M.S. \& Blades, J.C. 1997, ApJ 484, L125

\item
Silich, S.A., \& Tenorio-Tagle, G. 1998, MNRAS 299, 249

\item
Tinsley, B.M. 1980 Fund.Cosmic Phys. 5, 287

\item
Tolstoy, E. \& Saha, A. 1996, ApJ 462, 672

\item
Tosi, M., Focardi, M., Greggio, L., Marconi, G. 1989, {\it The Messenger},
 57, 57

\item
Tosi, M., Greggio, L., Marconi, G., Focardi, P. 1991, AJ 102, 951

\item
Tytler, D., O'Meara, J.M., Suzuki, N., Lubin, D., Burles, S. 2000 in
 The Light Elements and Their Evolution, IAU Symp.198, L. da Silva, M. Spite
 \& J.R. de Medeiros, p.125

\item
van Zee, L. 2001, AJ in press (astro-ph/0101135)
\end{description}
}

\end{document}